\input harvmac
\input psfig
\newcount\figno
\figno=0
\def\fig#1#2#3{
\par\begingroup\parindent=0pt\leftskip=1cm\rightskip=1cm\parindent=0pt
\global\advance\figno by 1
\midinsert
\epsfxsize=#3
\centerline{\epsfbox{#2}}
\vskip 12pt
{\bf Fig. \the\figno:} #1\par
\endinsert\endgroup\par
}
\def\figlabel#1{\xdef#1{\the\figno}}
\def\encadremath#1{\vbox{\hrule\hbox{\vrule\kern8pt\vbox{\kern8pt
\hbox{$\displaystyle #1$}\kern8pt}
\kern8pt\vrule}\hrule}}
\def\underarrow#1{\vbox{\ialign{##\crcr$\hfil\displaystyle
 {#1}\hfil$\crcr\noalign{\kern1pt\nointerlineskip}$\longrightarrow$\crcr}}}
%
\overfullrule=0pt

%

%

\font\zfont = cmss10 

\def\bigone{\hbox{1\kern -.23em {\rm l}}}
\def\ZZ{\hbox{\zfont Z\kern-.4emZ}}

\Title{hep-ph/9812208, IASSNS-HEP-98-102}
{\vbox{\centerline{NEW PERSPECTIVES}
\bigskip
\centerline{ IN THE QUEST FOR UNIFICATION}}}
\smallskip
\centerline{Edward Witten}
\smallskip
\centerline{\it School of Natural Sciences, Institute for Advanced Study}
\centerline{\it Olden Lane, Princeton, NJ 08540, USA}\bigskip

\medskip

\noindent
Synthesizing older ideas about the $1/N$ expansion in gauge theory,
the  quantum mechanics of black holes, and
quantum field theory in Anti de Sitter space,
a new correspondence between gauge theory and quantum
gravity has illuminated both subjects.
\Date{December, 1998}

I will not, obviously,
 be able in today's lecture to introduce all of the subjects that
will be covered in this school.\foot{This talk
was presented as the opening lecture at the 1998 School on Subnuclear
Physics at Erice.  It was based in part on the author's Klein Lecture
at Stockholm University.}   My somewhat more modest aim will be
 to explain some of the new advances in the quest for unification
 of the forces of nature.

But I have decided not to do this in the form of a standard review talk.
Instead, I thought I would explain the theoretical development of the
past year that seems most exciting to me.  This involves new ideas
that have combined together three longstanding ingredients,
namely:

(1) the $1/N$ explansion of gauge theories;

 (2) the thermodynamics of black holes;
 
 (3) quantum mechanics in Anti de Sitter spacetime.
 
 I will introduce these in turn and then describe the current synthesis.
 
 \bigskip\noindent{\it The $1/N$ Expansion}
 
 Some of the most basic problems in four-dimensional quantum gauge theories
 are still not really understood.  These include quark confinement,
 chiral symmetry breaking, and the mass gap in pure non-abelian
 gauge theories.
 
 We can exhibit confinement in computer simulations;  and we have
 (as  't Hooft will explain in a few days) a conceptual understanding
 of it based on electric-magnetic duality.  But we are not able in QCD
  to exhibit confinement  in any controlled 
 pencil and paper computation with well-defined approximations.   
 In short, we just do not have the sort of understanding of confinement
 that we aim to get for any physical phenomenon.
  Similar remarks apply for chiral symmetry breaking and the mass gap.
 As long as this situation persists, there is really no hope of
 computing hadron masses (and other strong interaction observables
 like hadron magnetic moments, scattering amplitudes, and the like)
 except by computer simulation.

 For a quarter of a century now \ref\thooft{G. 't Hooft, ``A Planar
 Diagram Model For Strong Interactions,'' Nucl. Phys. {\bf B72}
 (1974) 461.}, it has seemed that the best hope
of understanding  
the strong coupling aspects of QCD is via a $1/N$ expansion.
The idea is to replace the $SU(3)$ gauge group of strong interactions
by $SU(N)$ and expand in powers of $1/N$.  There are reasons
to believe that QCD simplifies dramatically for $N\to \infty$ and
that the $N=\infty$ theory is in fact, a free theory of hadrons.

If true, this is of fundamental importance; it means that taking $N$ large
separates the problem of formation of hadrons via mass generation and
quark confinement from the problem of the residual interactions of
hadrons.   These residual interactions lead to such complicated phenomena
(like nuclear physics) that there is no hope at all of ever getting
an exact solution of QCD.  Likewise, there is  no hope of ever 
understanding QCD  in a quantitative and controlled way unless
there is some gauge-invariant parameter that can be adjusted to turn
off the residual hadronic interactions while preserving the essential
mysteries that one wishes to explain.  It seems that $1/N$ is the parameter
that can play this role.

Moreover, there are hints from experiment that the real world is relatively
close to a limit in which confinement and the other nonperturbative 
phenomena are retained but the residual interactions are turned off.
For example, mesons, even relatively heavy ones, are comparatively
narrow (relative to their masses), suggesting some suppression of
the interactions leading to their decay relative to the strong interaction
that leads to their existence.  This suggests that the $1/N$ expansion
could be a reasonable quantitative as well as qualitative approach
to understanding the real world.  This really should not be so surprising;
fortune often smiles on approximations that are qualitatively correct,
and while $1/N=1/3$ is not so very small, it is not much larger
that the charge of the electron ($e=.303$, with $e^2/4\pi=1/137$).
Lattice simulations also seem to show that the $1/N$ expansion
is a good approximation at $N=3$, at least in three dimensions and
probably also in four \ref\teper{M. Teper, ``$SU(N)$ Gauge Theories
For All $N$ In 3 and 4 Dimensions,'' Phys. Lett. {\bf B397} (1997) 223, 
hep-lat/9701003, ``$SU(N)$ Gauge Theories In $2+1$ Dimensions,''
hep-lat/9804008.}.
For a review of some of the phenomenological arguments, see
\ref\witbar{E. Witten, ``Baryons in The $1/N$ Expansion,'' Nucl. Phys.
{\bf B160} (1979) 57.};
for more information on the $1/N$ expansion, see \ref\coleman{S. Coleman,
``$1/N$,'' in the proceedings of the 1979 Erice School on Subnuclear Physics}.

Going back to the early days of speculation about the $1/N$ expansion
in QCD, there have been reasons to believe that the solution of large $N$
QCD has something to do with a kind of string theory.  There are
at least three such reasons (which all largely go back to 't Hooft's
original paper): {\it (i)} QCD has strings; {\it (ii)} the
large $N$ Feynman diagrams are suggestive of strings; {\it (iii)}
phenomenology has apparently suggested a string description.

On the first point, QCD has strings, namely the strings or flux tubes
responsible
for confinement.  If one separates a quark from an antiquark, an electric
flux tube forms between them.  This gives a linearly growing energy, at least 
until the flux tube breaks!  Taking $N\to\infty$ separates the process
of formation of the flux tube, which we want to understand, from its
subsequent breaking and decay, which is also important but whose
study we would be happy to postpone until we understand why the flux
tube exists in the first place. The breaking of the flux tube
occurs by interactions that are
of order $1/\sqrt N$.  The confining flux tubes of QCD  behave like
strings at least macroscopically, and it is natural to at least
wonder if a more precise description of QCD can be given in terms of them.

The second point comes from the fact that the dominant
Feynman diagrams of the large $N$ limit are ``planar diagrams'' which
can be drawn on a sheet of paper with no lines crossing.  This
suggests the intuitive idea that nonperturbative effects will somehow
close up the holes in the Feynman diagrams, giving smooth string
worldsheets.  The strings would be interpreted, hopefully, as confining
flux tubes, giving an explanation of confinement.

The third point is that, after all, string theory was discovered almost
30 years ago because it seemed to describe some aspects of the strong
interactions correctly.  In hindsight, in this relation of string theory
to strong interactions, $1/N$ corresponds to the string coupling constant.
To the extent that the discovery of string theory was not a lucky historical
accident, it was discovered because of its analogy with the $1/N$
expansion of four-dimensional gauge theories; and this certainly encourages 
us to wonder how deep that analogy goes.

But unlike the $1/N$ expansion in some other models, the $1/N$ expansion
in gauge theories has been intractable for the last 25 years.  The
planar diagrams are simply too complicated to understand by any known
method.  The search for an understanding of the $1/N$ expansion
via a kind of string theory has  stimulated intense and at times
extremely fruitful work; a recent discussion is \ref\polyakov{A. M.
Polyakov, ``String Theory And Quark Confinement,''  hep-th/9711002.}.  
One important insight
has been that to describe gauge theory in four dimensions via a string
theory, the string theory must apparently be formulated in a world
of more than four dimensions, possibly by including an extra dimension
corresponding to Polyakov's Liouville field.

We now move on to our second subject.

\bigskip\noindent{\it Black Hole Thermodynamics}

What is a black hole?  

Classically, a black  hole absorbs and does not emit. Quantum
mechanically, this is impossible, since, for example, hermiticity
of the Hamiltonian implies that if there is a matrix element for
absorption, there must also be a matrix element for emission.

At this level, the paradox was resolved by Hawking in 1975
\ref\hawking{S. Hawking,  ``Particle Creation By Black Holes,''
Commun. Math. Phys. {\bf 43} (1975) 199.}.  
He showed that black holes do emit,
in a way that is thermal in the limit of a large black hole.
A black hole is characterized by a temperature, which for a Schwarzschild
black hole in four dimensions of mass $M$ is of order
$T\sim 1/G_NM$ (with $G_N$ being Newton's constant), and has an
entropy $S=A/4G_N$, with $A$ the area of the event horizon.
(Such thermodynamic properties had been conjectured earlier
by Bekenstein based on classical arguments.)

For a solar mass black hole, this is a fantastically small temperature
and a huge entropy.  For example, the entropy of a solar mass black
hole is vastly bigger than the entropy of the sun in its present state.

What does the entropy of a black hole really mean?  In the rest of physics,
the entropy is the logarithm of the number of quantum states.
But what are the quantum states of a black hole?  Evidently there
are many of them, as the entropy is so large.   If the entropy
of a black hole really counts the quantum states, it must be that a quantum
black hole is not {\it at all} fully specified by giving just its
mass, charge, and angular momentum -- which give a complete specification
classically.

Identifying the quantum states of a black hole is a question that combines
quantum mechanics and gravity.  So -- at least among theories we know
now -- this question is really only well-posed in string theory, which
is the only concrete candidate we have for a consistent quantum theory
that incorporates General Relativity as a limiting approximation.
But string theory in its first quarter century was not sufficiently
well understood to shed any light on the quantum nature of black holes.
The first partial answers have begun to emerge only in the last
few years.

The idea is to consider black holes that are built out of ``$D$-branes.''
$D$-branes are nonperturbative excitations of string theory that have
a strange ``matrix'' property.  
A single $D$ brane in $n$ dimensions
has position coordinates $x_1,x_2,\dots, x_n$, just like any other
particle.  But for a system of $N$ identical $D$-branes, the
$x_i$ become $N\times N$ matrices that in general do not commute.
The situation is a bit like the quantum mechanical noncommutativity
of position and momentum -- the familiar $[p,x]=-i\hbar$ -- but now
it is the different components of the position that do not commute.
Moreover, there is a $U(N)$ symmetry acting on the $N\times N$ position
matrices.  It fact, this $U(N)$ symmetry is a {\it gauge symmetry}.

If we make a black hole from $N$ $D$-branes, we get a $U(N)$ gauge
theory that describes certain black holes.  To be more exact, these
objects can be described as semiclassical black holes, with a horizon
size large compared to the Planck length, only if $N$ is large
(and certain other conditions are obeyed).  So to compare to the
thermodynamic description of black holes, we must take $N$ large.

For certain black holes, with only relatively ``easy'' gauge theory
results, it has proved possible \ref\strom{A. Strominger and C. Vafa,
``Microscopic Origin Of The Bekenstein-Hawking Entropy,'' Phys. Lett. {\bf B379}
(1996) 99.}
to understand and count the quantum states,
making contact with the Bekenstein-Hawking formula for the entropy.
This success is limited to only certain kinds of black holes, and it does not
address deeper questions like whether quantum mechanical unitarity
is preserved in the formation and evaporation of a black hole. 

But it raises the question: Would deeper results about black holes
be related to more difficult aspects of large $N$ gauge theory?

To explain as much as we know of the answer to this question, I must
move on to our third subject.

\bigskip\noindent{\it Quantum Mechanics In Anti de Sitter Space}

Anti de Sitter space is a maximally symmetric world with negative
cosmological constant.  In one convenient coordinate system, the
metric looks like
\eqn\loooks{ds^2={dr^2\over 1+r^2}+r^2d\Omega^2-(1+r^2)dt^2.}
This spacetime has many peculiar properties which are related to the fact
that the coefficient of $dt^2$ grows as $r\to\infty$.

One very desireable consequence of this is that it gives
\ref\hawpage{S. Hawking and D. Page, ``Thermodynamics of Black Holes
in Anti-de Sitter Space,'' Commun. Math. Phys. {\bf 87}
(1983) 577.} an elegant infrared cutoff
in the thermodynamics of black holes.  Black hole thermodynamics
suffers, apart from everything else, from the instability of the usual
thermal ensemble in the presence of gravity.  To speak of the temperature
of a black hole can only be precise if the black hole is in thermal
equilibrium; but in Minkowski space, a thermal ensemble of any positive
temperature is unstable against gravitational collapse (which will
produce a black hole much bigger and heavier than the one we originally
undertook to study).  The factor of $1+r^2$ in the Anti de Sitter
metric gives a spatial dependence to the effective temperature,
with the consequence that a thermal ensemble in Anti de Sitter space
is perfectly stable.
In particular one can really exhibit a black hole in thermal
equilibrium with radiation in Anti de Sitter space, if the mass of the
black hole is large enough.

This is a convenient context for studying quantum black holes; but what
do we want to do with them?  

At least with a little bit of hindsight, one answer to this question is:
We would like to explore the notion of ``holography.''
To make sense of the puzzles of quantum black holes, it
has been argued by 't Hooft \ref\thol{G. 't Hooft, ``Dimensional Reduction
In Quantum Gravity,'' in {\it Salamfest 1993}, p. 284, gr-qc/9310026.}
and Susskind \ref\sussk{L. Susskind, ``The World As A Hologram,''
J. Math. Phys. {\bf 36} (1995) 6377.} (and by Thorn \ref\thorn{C. Thorn,
``Reformulating String Theory With The $1/N$ Expansion,'' lecture at first
A. D. Sakharov Conference on Physics, hep-th/9405069.}
in the string context)
that nature should be ``holographic,'' a property that one might well
characterize as magic.  The idea of holography is that, in contrast
to the usual  description of physics by degrees of freedom
that are approximately local, a theory with quantum gravity should
have a description by degrees of freedom that are defined on the boundary
of space.  

This would make it possible to describe
 the formation and evaporation of a black hole just
in terms of things that happen at spatial infinity, without delving into
the details of what happens near the black hole horizon.  Thus,
a holographic description would make unitary in black hole physics
manifest.
But to describe what happens in the ``bulk'' of spacetime by dynamical
variables that ``live'' at infinity
 seems well-nigh impossible, given the apparent locality
of physics.  In classical field theory, it probably really is impossible.
So holography, if it holds, must be achieved in a way that 
does not commute with the passage to a classical limit.

The same factor of $1+r^2$ that makes Anti de Sitter space an attractive
arena for studying black holes
has other consequences that at first sight
seem rather perplexing.  A small calculation, given the metric as written
above, shows that spatial infinity (that is, $r=\infty$) is at an infinite
distance if one tries to get there along a spacelike path.  But along
a lightlike path, things are different.  A light ray can reach infinity
in Anti de Sitter space with only a finite time (or affine parameter) elapsed;
so a signal propagated at the speed of light can reach the end of Anti
de Sitter space and return in a finite time.  Hence, to make sense of
quantum theory in Anti de Sitter space, one needs a boundary condition
at spatial infinity.  This caused perplexity at first, but it was
eventually learned \ref\avis{S. J. Avis, C.
Isham, and D. Storey, ``Quantum Field Theory In Anti-de Sitter
Space-Time,'' Phys. Rev. {\bf D18} (1978) 3565.} that one can put
such a boundary condition and get a well-defined quantum mechanics
in Anti de Sitter space.  The relation between the bulk of Anti de Sitter
space and its boundary has fascinated physicists for a long time,
going back to early work by Dirac \ref\singletons{For example, see
M. Flato and C. Fronsdal, ``Quantum Field Theory of Singletons: The Rac,''
J. Math. Phys. {\bf 22} (1981) 1100.}.

\bigskip\noindent
{\it Synthesis}

It is these ingredients that have now been combined, in a daring
conjecture by Maldacena \ref\malda{J. Maldacena, ``The Large $N$ Limit
Of Superconformal Field Theories And Supergravity,'' hep-th/9711200.}, 
subsequently
formulated in a more precise form \nref\kleb{S. Gubser, I. R. Klebanov,
and A. M. Polyakov, ``Gauge Theory Correlators From Non-Critical
String Theory,'' hep-th/9802109.}
\nref\uwit{E. Witten, ``Anti de Sitter Space
And Holography,'' hep-th/9802150.} \refs{\kleb,\uwit}.  
The key, assuming that we want
to describe black holes in four dimensions, is to start with Anti de Sitter
space in {\it five} dimensions.  The symmetry group of five-dimensional
Anti de Sitter space is $SO(2,4)$.  Now consider Minkowski space in
four dimensions.  Its {\it conformal } symmetry group is the same
group $SO(2,4)$.  

So there are two kinds of physical theories with $SO(2,4)$ symmetry:
any relativistic theory at all in five-dimensional Anti de Sitter space;
or any conformal field theory in four-dimensional Minkowski space.

Now, Anti de Sitter space of five dimensions has a sort of boundary
at spatial infinity.  As I explained above, light rays can reach the
boundary in finite time, and for this reason a boundary condition at
spatial infinity is needed in order to make sense of quantum field theory
in Anti de Sitter space.  The boundary of Anti de Sitter five-space is
a four-manifold which is in fact a copy of four-dimensional Minkowski space
(conformally completed), and from this point of view the $SO(2,4)$
symmetry of Minkowski four-space is simply the restriction to the boundary
of the $SO(2,4)$ symmetry of Anti de Sitter five-space.

The new insight in the recent work is that a {\it gauge theory}
(or more generally a conventional flat space quantum field theory
without gravity) on Minkowski four-space can be equivalent to a {\it theory
with quantum gravity} in Anti de Sitter five-space.  (Likewise, a conventional
theory on Minkowski $n$-space can be related to a theory with
quantum gravity on Anti de Sitter $n+1$-space.)  

This correspondence
would hold for any quantum gravity theory in a spacetime that is
asymptotic to Anti de Sitter space.  The only known candidates
come from string theory.  In the string theory case, we can in many
cases determine just which conformally invariant gauge theory on Minkowski
space is related to a given theory in Anti de Sitter space.
On the gauge theory side, because $SO(2,4)$ acts on Minkowski space
by conformal transformations, the gauge theory that arises
 is necessarily conformally invariant (its beta function vanishes);
an example, which plays an important role in the correspondence,
is the ${\cal N}=4$ super Yang-Mills theory.

The correspondence between these two types of theories
is ``holographic'' in that a gravitational theory in the
``bulk'' of Anti de Sitter space has an alternative description in which
all the degrees of freedom ``live'' on the boundary at spatial  infinity.
The semiclassical (weak coupling or $\hbar \to 0$) limit of the gravity
theory becomes a large $N$ limit of the gauge theory, so as expected
holography does not commute with the passage to a classical limit.

The existence of this holographic description of the supergravity/string theory
makes quantum mechanical unitarity manifest.  Black hole entropy is mapped
to the entropy of the gauge theory, and the fact that the black hole
entropy is proportional to the surface area of the horizon becomes
a standard scaling relation of conformal field theory.  (But, frustratingly,
the constant of proportionality can with our present understanding
be computed only in the $2+1$-dimensional
case.)  

Large $N$ is reinterpreted as a semiclassical or weak coupling
limit --
at least for the specific gauge theories that arise in the correspondence
with Anti de Sitter space -- rather as was anticipated 25 years ago.

After suitable perturbations, this correspondence can be used to deduce
properties of more realistic gauge theories that are not conformally
invariant.
For certain four-dimensional
gauge theories, which we conjecture but cannot prove to be in the
same universality class as the pure gauge theory,
confinement and the mass gap are interpreted as consequences of
the topology of Euclidean black holes.
In this framework, the deconfinement phase transition for gauge theories 
is mapped to a phase transition for Anti de Sitter black  holes found
long ago by Hawking and Page \hawpage.  The endstage of evaporation of
a Schwarzschild black hole -- where one must face the unresolved puzzles
of quantum black hole physics -- is mapped to the decay of the high temperature
phase of the gauge theory, after being supercooled to temperatures at
which it is not thermodynamically favored. 

One can further identify the Wilson and 't Hooft loop operators and the
``baryon vertex'' of gauge theories in terms of structures that appear
in string theory on Anti de Sitter space.  Many of their properties
can be identified in this framework.

\bigskip\noindent{\it Outlook}

What can one do, and what cannot one do, with the present understanding
of this subject?

One can use known properties of gauge theory to get important
qualitative insights about gravity, and vice-versa.
But many questions that one would most like to answer remain out
of reach.  One cannot understand the endpoint
of black hole evaporation, because in the relevant regime, the
gauge theory is strongly coupled.  And one cannot compute the hadron
masses in four-dimensional pure gauge theory because in the relevant
regime, the supergravity approximation to the Anti de Sitter space theory
breaks down.  

So I conclude with the same question that we asked before:
Would (yet) deeper aspects of the $1/N$ expansion of gauge theories
be related to (yet) deeper aspects of quantum black holes?

\bigskip
Research supported in part by NSF Grant PHY-9513835.  
\listrefs
\end